# Social Business Transformation through Gamification

Jitendra Maan

Tata Consultancy Services Ltd.,
*jitendra.maan@tcs.com*

## *Abstract*

*Being an emerging business practice, gamification is going to the mainstream to enable and transform social business initiatives across enterprises. With the consistent focus on customer behavior and experience, there is a paradigm shift in thinking about how Gamification and Social initiatives together help to increase the engagement level of knowledge worker, yielding better business results. Business scenarios for gamification are wide spread ranging from customer service and support to communities and collaboration.*

*The Paper discusses the characteristics & mechanism to learn from games that are important for businesses to understand and apply. It also gives insights on gamification trends, real-world business challenges and also describes on how game thinking can revolutionize the business and create an engaging experience.*

## 1   Introduction

Gamification has become a modern business practice that uses game mechanics and game design elements to measure, influence and reward target user behaviors. It takes the essence of the game characteristics like – goals, rules, playfulness, elements of fun, feedback, reward and promotions - applies them to solve the real-world business problem. These game mechanics when applied in the non-gaming context, work as a catalyst for making technology more engaging by influencing user behavior and social interaction methods.

Gamification allows enterprise to glean valuable insights into customer, employee behavior and activities across various touch points, including website, mobile applications and social collaboration applications. Companies will use this data to determine what content and experiences are high-value and see how customer actions correlate with business success. With gamification, they'll be able to clearly see the employee behaviors across applications and truly understand what motivates them to further engage with right gaming mechanics. Fundamentally, Gamification acts as a layer on the top of Social Collaboration software to gain valuable insights into customer behavior, opinions and employee activities across various touch points.

## 2   Key Trends – Gamification and Social Engineering

Gamification and Social Engineering, together has tremendous potential in engaging with stakeholders - Customers, Employees and Partners. Some of the key gamification trends that will definitely gain a lot of traction among businesses as below –

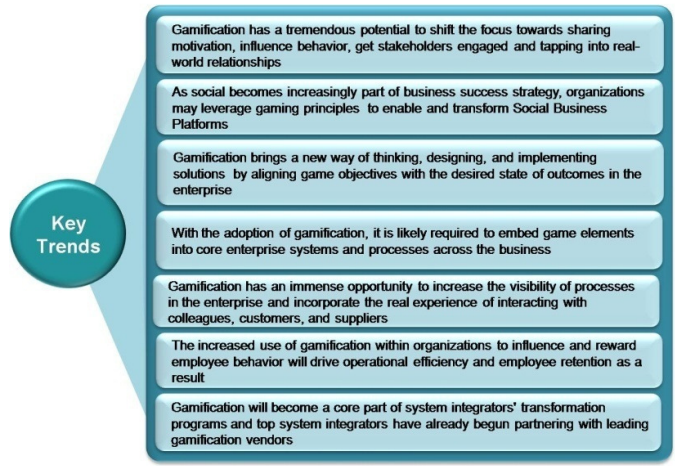

**Fig. 1.** Enterprise Gamification trends

## 3   Real-world Business Challenges

Some of the real-world business challenges that become major roadblock to gamification adoption across enterprises as given below:

- **Culture Gap** - Gamification as a business discipline is evolving with the current usage of gamified mechanics but the ultimate success depends on the culture change in the organization which is the major barrier to its adoption in businesses.
- **Lack of 'Win States' or 'Success Metrics'** – What target behaviors and success metrics will define if gamified system is a success or failure
- **Lack of Motivation –** No value derived from encouraging behavior to make boring activities interesting
- **Lack of understanding the needs of the game player** like Demographics (age, gender etc) and Psychographics (Personalities and their values)
- **No meaningful choices available to the user** – Target activities and behaviors are not sufficiently engaging
- No structure laid out to model the target behaviors
- Lack of fun element de-motivates the users to engage with the application/system
- Lack of right gamification tools, platforms and multi-channel support

## 4   Gamification and Game Thinking

Games have already been an integral part of our society where people appreciate the feeling of earning points, rewards and autonomy by overcoming challenges and obstacles with an element of fun. According to leading experts and market thinkers, design thinking is a process that all business engages for specific purpose.

Gamification applies the dynamics and mechanics of psychology that make games so engaging and sticky. It is the means to provide information into the system and facilitating the process that engages that kind of sharing. Gamification typically makes more sense when game design thinking is applied to non-gaming applications like enterprise business applications, collaboration and communication application suites etc.

10

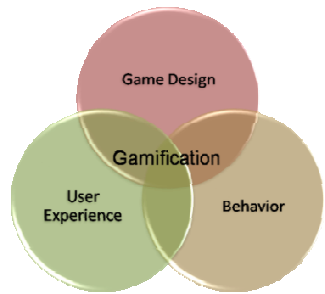

**Fig. 2.** Gamification Overview

In above figure, gamification is a factor of the following –
- **Game Design** – The way application needs logic, games are also designed considering an internal logic in mind. In case, games do not allow performing something logical without any warning, it fails to engage the user for a long.
- **Behavior** – High-end user engagement to influence the target behaviors whereas, engagement level depends on the type of the game players –
    ─ **Achievers** who are high focused on game-oriented goals
    ─ **Explorers** who are innovative and find the hidden part of the game
    ─ **Socializers** mainly focused on engaging, sharing, collaborating information
    ─ **Killers** who always want to create trouble/problems for other participants
- **User experience** – Most interaction design is about process and knowledge worker efficiency where "Gamefulness" and "Playfulness" attributes are important in user experience design.

While designing the gamified applications, designer makes a due consideration of the game mechanics to create the right user experience.

In the below diagram, some of the important characteristics of the gamified design as given below –
- **Goals –** Organizations define goals for business users and monitor their behavior and reward them on their achievements in the overall journey. Design of the gamification system need to constantly refer back to achieving those goals. Games lose their mettle in case they are not tied to the goals.

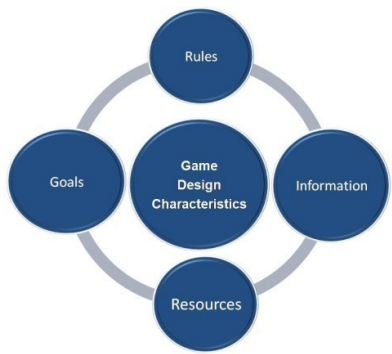

**Fig. 3.** Key characteristics of Game Design

- **Rules –** Rules become an inherent part of the game. Each game has a series of rules provided by game designer which are applied by the player. In business context, rules are often captured in the form of wire-frames.

11

- **Information** – Games provide player the most relevant information to make decisions appropriately. At every stage, players review the state of the game and make a decision about what to do next. It is the responsibility of the game designer to ensure that players are provided relevant information to manage their resources.

- **Manage resources in terms of time, effort and Status** – Say for example, Blogger experience on the websites and portal shows the level of effort that is required to complete certain tasks. Progress bar for installing and configuring products, tools indicates time for the activity/ task. Several social media sites give a status or progress bar to encourage user to continue and complete their social profiles.

## 5 Gamified Design – Key Elements

Gamification has several use cases for successful application in social media. It has now become an innovative way to engage and motivate customers using game mechanics. Several organizations have started creating social loyalty initiatives to improve their brand image. In order to achieve this, they are leveraging gamification behavior platform to structure such strategies.

The typical architecture of any gamification initiative is essentially based on key game elements including the following –

### 5.1 Rewards & Incentives

To stay competitive, organizations prefer to run reward campaigns to offer discounts, promotions and incentives to their employees, customers and partners through Loyalty programs. Organizations design rewards structure that encourages desired behaviors in employee-facing environment. Some organizations prefer to design two level reward mechanisms - **Badges** for adoption and frequency of use on particular platform and **Karma points** for desired action within the platform

By virtue of fact, most organizations often use extrinsic incentives to motivate their employees which, at times fail to intrinsically motivate them to work. Such external rewards are good if they are aligned with internal psychological need of the user. On practical ground, it is much more effective to practice intrinsic motivators like social competition and continuous feedback highlighting achievements and mastery levels.

### 5.2 Badges

Gamification techniques when embedded into social applications include badges demonstrating different level of achievements when participation milestones are reached.

### 5.3 Leaderboards

It is an emerging practice in forward-looking organizations to assign Leaderboards in different areas of domain expertise across business functions. People normally like to validate if they are performing well as per expectations or not. Leaderboard helps people to know where they stand relative to their colleagues or peers thereby inculcating a spirit of competition.

### 5.4 Point System and Scores

Besides rewards and incentives, there are other viable ways to motivate and encourage the desired behaviors by earning points and further chance to win the awards and incentives. Game player earn points based on level of participation. Points come in many different forms like redeemable points, Skill, karma and Reputation etc.

The criteria for awarding Points is broadly depend on following key attributes –Speed of Response, Frequency of Participation, Quality of participation and Learning Continuum. Gamification examples against each of these attribute is shown in below diagram –

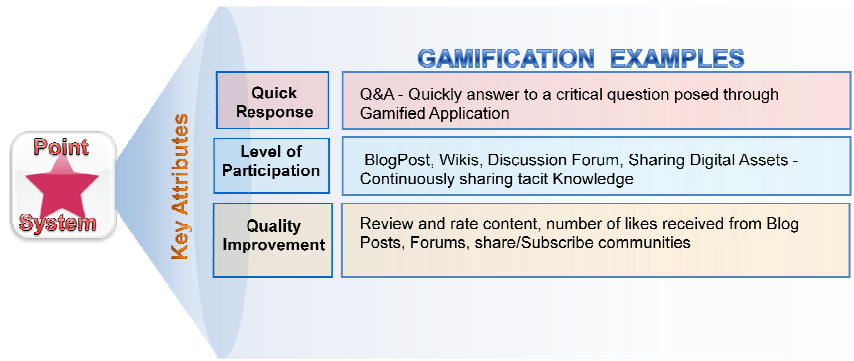

**Fig.** 4**.** Key Criteria for awarding points in a gamified Application

### 5.5 Competition

Competition is something that can describe a situation where success can be measured in terms of outcome. Competition may take one or other form of several dimensions including speed, accuracy, creativity, strategic tactics, Knowledge and time, for instance football game requires physical elements (Strength, speed and accuracy) along with mental tactics and knowledge of the opponent team. Although, games are mainly characterized by competition but Game designers usually focus on creating a team-like collaboration environment.

### 5.6 Social Connection

Social Connections leverage social networks to create competition and provide customer support. With the high penetration of mobile web and high adoption of mobile devices and tablets, social networks may provide instant access to social connections anytime anywhere which increases the level of engagement and interactions.

### 5.7 Levels & Reputation

It signifies the level of user engagement across the business value chain which becomes a basis for awarding the players once they reach a specified level.

A user generates reputation when he gets an enough attention to the questions and answers posted by him. Reputation is the clear measure of the trust build in the community and gives an understanding of relevance of your questions and answers in right context.

## 6 Enterprise Gamification

Mostly enterprise applications may fall into one of the two gamification buckets – Internal and External Gamification where employee-facing application qualifies for internal gamification and the typical customer-facing scenarios may better fit into external gamification category. Enterprise gamification takes learning from computer games and then integrates them in the business process to influence behaviors.

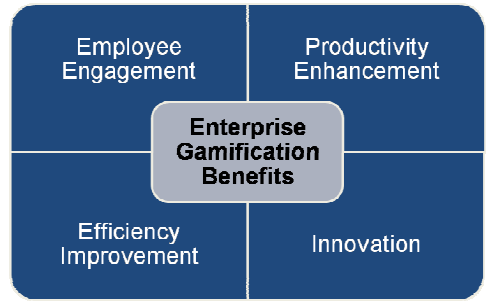

**Fig. 5.** Enterprise Gamification Benefits

The benefits of Enterprise Gamification are broadly categorized into four areas–

### 6.1   Employee Engagement

Many large enterprises are attempting of leverage gamification to encourage their employees to make valuable contribution to their existing collaboration and communication platforms. Most of them are experimenting behavior platform in such a way that it is used to build game mechanics to recognize key contributors and to design user-centric motivations and achievements to reward user behaviors across enterprise-wide community networks.

From game designer perspectives, it is not easy to design game mechanics to make boring activities as part of a game to improve engagement and overall adoption, for example, a sales person gets points on creating a new opportunity in CRM system and such earned points further motivates him to earn more points in completing his activities in CRM system. In this journey, sales person get recognition through badges, special titles and he progressed to mastery level and become a face on Leaderboard along with his titles and rewards. Overall, the sales team will get more accurate picture of the opportunities and in a typical sales cycle, these teams are motivated to improve the quality of their services to the customer.

### 6.2   Productivity Enhancement

Most of the product companies look at gamification as a core to their product strategy from the very beginning to revamp their existing branding sites by enabling –

- Real-time rewards for user behaviors,
- Leaderboards for significant achievements, and
- A natively gamified user experience

In today's competitive environment, organizations are looking at Key Performance Indicator (KPIs) to integrate them in their business metrics where data flows from gamification engine that allows them to monitor employee performance and it becomes a key decision driver to improve efficiency and overall productivity. For example, in a typical Call center scenario, each agent performance is judged by the fact how quickly he deals with the customer call where call operators foresee to look to gamification as a technique to help them manage these customer service engines better. Customer satisfaction numbers along with virtual rewards are shown on dashboards that give them real-time feedback.

### 6.3  Efficiency Improvement

Efficiency as a term, just make people work better by doing everything they can do efficiently. Specifically the example that I am going to talk about email as most of the critical time for productive work for business users is spent in email management, it is not their job primarily to answer/response to every email, but they have to do it as part of their job, people waste tons of time in reading and answering such emails, clearing their mailbox so and so forth. Gamification practices are being used to improve the efficiency of such processes, by embedding game mechanics that you can see every moment to check how much time you spent in emails, progress bar telling how much you are left with pending mails along with a point system for quickly disposing off your emails, for deleting or responding to in the way that you can close these tasks in your mail box. It encourages you to deal with email efficiently, saving lot of time.

### 6.4  Innovation

Innovation is a center-stage theme for global organizations to stay competitive and it is imperative for them to continuously encourage their employees to come up with great ideas and thoughts and realize them to provide business benefits. Enterprises are now leaving no stone un-turned in capturing and realizing the full potential from the effective use of social platforms and technologies under overall transformational umbrella where new services, process innovation and creative product ideas will drive the market. Several organizations are leveraging gamification mechanics to drive such initiatives, say for example, concept like Idea marketplace similar on the lines of virtual stock market helps encourage all stakeholders to create a rewards and motivation platform to incubate, share and execute ideas from all business domains.

## 7  Leading Enterprise Gamification Platforms

Gamification Platform vendor landscape is growing rapidly. A number of vendors have already developed their own Gamification Platforms, ready to be customized to specific organization needs. Following table summarizes the list of major platforms –

| Category | Gamification Platform (s) |
| --- | --- |
| Enterprise | BadgeVille, BunchBall, Actionable, BigDoor, Gamify |
| Social | CrowdFactory, Socialtype, Gigya, Badgy, Gamification |
| CRM | Fusion, Hoopla |
| Loyalty | CrowdTwist, BulbStorm |
| Support & Services | Nicely |
| HCM | MindTickle |

**Fig. 2.** Leading Gamification Platforms

## 8  Future of Gamification

Today, most of the global organizations have been on the leading edge of adopting communities powered by nexus of technologies – Mobile, Cloud, Social, and Business Intelligence for their extended ecosystem of customers, partners, employees, suppliers and industry domain experts. However, as more and more enterprises will adopt gamification, business user expectations will change over time, but there would always be the need for a strong gamification community to start picking up the pace from internal collaboration and innovation. Gamification Frameworks will play an important role to offer design models for customer and employee-facing initiatives for a sophisticated and fast gamification deployment.

## 9 Concluding Thoughts

Gamification has shown tremendous growth in achieving various social business initiatives in the organization and it brings new way of thinking by aligning game objectives with the desired outcome in the organization. Today, Social business strategies are based more on high-end user engagement and connections where behavior patterns are highly dynamic.

Most of the global organizations have assessed the need to build self-sufficient customer communities that would spur innovation, capture voice of customer analytics, and streamline services and support functions. Here, gamification behavior platforms and reputation engines are becoming key drivers to recognize users who have performed key behaviors and motivate them to actively engage within these communities. Over the next few years, gamification will become a core part of system integrators' transformation programs.

There is a big opportunity for next-generation organizations to explore new game practices to understand building blocks of enterprise gamification and such organizations would harness gaming principles to influence behaviors of key stakeholders in the eco-system. In a typical social mobile scenario, many organizations have successfully leveraged gamification by incorporating social context and location-aware services. Enterprises would be forced to explore new game practices and attributes to develop gaming strategies to drive game-like behavior in existing applications/tasks to make them more engaging to the end-users.